\documentclass{article}

\usepackage{PRIMEarxiv}

\usepackage[utf8]{inputenc} 
\usepackage[T1]{fontenc}    
\usepackage{hyperref}       
\usepackage{url}            
\usepackage{booktabs}       
\usepackage{amsfonts}       
\usepackage{nicefrac}       
\usepackage{microtype}      
\usepackage{lipsum}
\usepackage{xcolor}
\usepackage{subfig}
\usepackage{tabularx}
\usepackage{multirow}
\usepackage{enumitem}
\usepackage{cite}
\usepackage{fancyhdr}       
\usepackage{graphicx}       
\graphicspath{{media/}}     

\title{Towards Reconciling Usability and Usefulness of Explainable AI Methodologies
}

\author{
  Pradyumna Tambwekar \\
  School of Interactive Computing \\
  Georgia Institute of Technology \\
  Atlanta, GA\\
  \texttt{pradyumna.tambwekar@gatech.edu} \\
   \And
  Matthew Gombolay \\
  School of Interactive Computing \\
  Georgia Institute of Technology \\
  Atlanta, GA\\
  \texttt{matthew.gombolay@cc.gatech.edu} \\
}

\begin{document}
\maketitle

\begin{abstract}
Interactive Artificial Intelligence (AI) agents are becoming increasingly prevalent in society. However, application of such systems without understanding them can be problematic. Black-box AI systems can lead to liability and accountability issues when they produce an incorrect decision. Explainable AI (XAI) seeks to bridge the knowledge gap, between developers and end-users, by offering insights into how an AI algorithm functions. Many modern algorithms focus on making the AI model ``transparent'', i.e. unveil the inherent functionality of the agent in a simpler format. However, these approaches do not cater to end-users of these systems, as users may not possess the requisite knowledge to understand these explanations in a reasonable amount of time. Therefore, to be able to develop suitable XAI methods, we need to understand the factors which influence subjective perception and objective usability. In this paper, we present a novel user-study which studies four differing XAI modalities commonly employed in prior work for explaining AI behavior, i.e. Decision Trees, Text, Programs. We study these XAI modalities in the context of explaining the actions of a self-driving car on a highway, as driving is an easily understandable real-world task and self-driving cars is a keen area of interest within the AI community. Our findings highlight internal consistency issues wherein participants perceived language explanations to be significantly more usable, however participants were better able to objectively understand the decision making process of the car through a decision tree explanation. Our work also provides further evidence of importance of integrating user-specific and situational criteria into the design of XAI systems. Our findings show that factors such as computer science experience, and watching the car succeed or fail can impact the perception and usefulness of the explanation. 
\end{abstract}


\section{Introduction}

As the breadth of applications within Artificial Intelligence (AI) grows, there is an widening chasm between developers of AI systems and consumers that these systems should cater to. 
This rift between end-user and technology leads to a decrease in trust and satisfaction in autonomous systems~\cite{matthews2019individual}. Humans understandably become suspicious towards these systems and are less tolerant to failures and mistakes~\cite{das2021explainable, kwon2018expressing, robinette2017effect}.
Explainable AI (XAI) was thus proposed as a means for developers to engender greater confidence in these systems by enabling users to understand the inner-workings and decision making process of AI algorithms~\cite{xu2019explainable, jacovi2021formalizing}. 
As such, XAI systems have now been broadly deployed in various capacities such as for banking~\cite{grath2018interpretable}, healthcare~\cite{pawar2020explainable}, robotics~\cite{anjomshoae2019explainable} among other domains. 

\begin{figure}
\centering
  \includegraphics[width=0.7\textwidth]{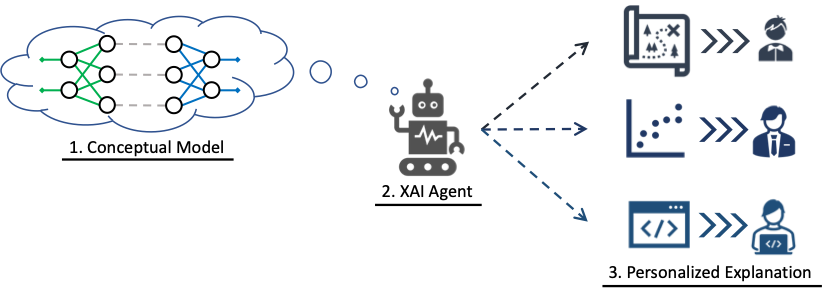}
  \caption{The job of an XAI agent is to present the user with an explanation they can apply towards understanding how the AI agent works. However, every agent has their own individual needs and context which the agent needs to cater to. A banker may want to be presented an explanation in numbers, whereas a student may prefer a more visual explanation. To effectively satisfy stakeholders, the AI agent needs to understand who the user is and formulate an explanation specifically suited to them. 
  }
  \label{fig:teaser}
\end{figure}

However, the increase in technological demand of these systems lead to a rise in predominantly ``model-centric'' explainable AI solutions. 
These approaches tackle the problem of XAI by ``opening-up'' the black box of deep neural networks through post-hoc or transparency approaches~\cite{xu2019explainable}.
Transparency-based approaches seek to expose the internal mechanisms of an algorithm in a simpler format. Prior work has established strong baselines for transparency through gradient-based relevance methods~\cite{samek2017explainable, simonyan2013deep}, or presented inherently interpretable architectures wherein the explainability comes from the nature of the architecture~\cite{humbird2018deep, silva2020neuralencoding, letham2015interpretable}. 
Contrarily, post-hoc approaches provide additional context to a user on an instance- or behavior-wise basis to explain an action or an output of a system. Such approaches include attention visualization~\cite{ghaeini2018interpreting}, behavior summarization~\cite{amir2019summarizing, lage2019exploring}, model reconciliation~\cite{chakraborti2017balancing}, and much more. 


Unfortunately, these methods gloss over a fundamental component of the interaction experience, i.e. the individual~\cite{wang2019designing}. Each user who may need to work with such a system has a unique set of needs, and will operate in a specific situational setting. To effectively cater to our stakeholders, we need better mechanisms to understand their needs and socio-technical dispositions and study the ways these factors influence their ability to effectively utilize an explanation~\cite{ehsan2020human, ehsan2021expanding, matthews2019individual}. 
XAI is not a one-size-fits-all problem and the most ``correct'' explanation is not always the most understandable~\cite{zhou2022exsum}. 
Ill-fitting XAI tools can have a regressive effect on AI safety by creating a false sense of security, as incomprehensible explanations do not enable users to accurately identify and diagnose potential points of failure~\cite{ghassemi2021false}.
To develop accessible XAI methodologies we need to understand needs of a user to develop explanations that suit them~\cite{liao2020questioning}. 

In this paper, we present a user study in which we compare multiple modalities of explanations with regards to usability and acceptance. Our goal is not to find the ``best'' XAI modality. Rather, we aim to apply these disparate modalities to better understand the dispositional factors which may influence an individuals preference towards an explanation. We conduct a novel human-grounded evaluation in which we studied which modality of explanation is the most helpful for a user in understanding the decision making process of a car on a highway. We developed a forward simulation protocol in which we tested a participant's ability to interpret four modalities of explanations for a self-driving car on a highway.  
Our work also seeks to unpack the relationship between perceived usefulness of an explanation and actual usefulness, which we define as how well a participant is able to apply the explanation towards predicting the behavior of the AI agent. 
Perception of usability influences adoption of the agent, whereas objective understandability dictates how beneficial the explanation will be to the stakeholder. 
Through our qualitative and quantitative analysis, of subjective and objective metrics of explanation usefulness, we seek to present a better understanding of how to connect users to the right explanation which suits their individual context and characteristics.  

We do not claim that we are the first to highlight the need for personalized explainability mechanisms. However, we provide support to contemporary work studying this topic~\cite{conati2021toward, millecamp2020s, silva2022explainable}, by incorporating an objective metric of simulatability in addition to subjective usability, via the Technology Acceptance Model (TAM)~\cite{davis1989perceived}, to understand the suitability of an explanation to a user. 
We identify key demographic factors, that elucidate which XAI method is more helpful for individual users. 
Our analysis, highlights issues regarding a lack of internal evaluative consistancy of XAI modalities, by demonstrating instances where users objectively better understand the underlying working of the self-driving car with the help of an explanation, but subjectively prefer a different modality because the first explanation was ill-fitting towards their distinct disposition~\cite{zhou2022exsum}. 
Finally, unlike prior work, we conduct our study within an autonomous driving domain. 
Driving is an accessible domain, as most people either possess driving experience or have been passengers in cars, making it an easily-understandable task for participants of our study.
Autonomous driving is a safety-critical application and thus is a highly relevant domain for explainability due to the ethical and liability concerns involved~\cite{stilgoe2019self, zablocki2021explainability}. 
Therefore, it is important to better understand the factors that influence the perception and ability to apply the explanations in these scenarios.  
To summarize, our contributions are as follows, 
\begin{enumerate}
    \item We present a novel study design to compare multiple modalities of explanations through both subjective metrics of usability and acceptance as well as objective metrics of simulatability.
    \item We conduct qualitative and quantitative analysis on data from 231 participants to elucidate individual preferences of explanation modality, as well as highlight the effect of situational or dispositional factors on the perception of the XAI agent.
    \item Our results highlight a lack of consistency in evaluative preference of explanation modalities, by showing that although participants rated text-based explanations to be significantly more usable than the decision tree explanation, the decision tree explanation was found to be significantly more useful for simulating the functionality of the self-driving car. 
\end{enumerate}

\section{Related Work}
\subsection{Explainable AI methodologies}
Explainable AI is a prominent area of research within artificial intelligence. 
The most prevalent explainability methods are model-based approaches, which seek to explain the black-box of a deep neural network. 
A popular preliminary approach was by visualizing the outputs and gradients of a deep neural network~\cite{yosinski2015understanding, simonyan2013deep, ghaeini2018interpreting, selvaraju2017grad}. 
These methods provided informative visualizations of neural network outputs and parameters in order to enable users to interpret the functionality of the network. 
However, it has been found that approaches that rely on visual assessment can sometimes be misleading, as they may be specific to unique data or modelling conditions, and can be highly susceptible to outlying outputs that contradict the explanation ~\cite{adebayo2018sanity, serrano2019attention, kindermans2019reliability}. 
Prior work has also sought to transform uninterpretable deep networks into interpretable architectures or modalities such as decision trees~\cite{humbird2018deep, silva2020neuralencoding, paleja2021utility}, or bayesian rule lists~\cite{letham2015interpretable}, and generate explanations by exploiting the ``white-box'' nature of these architectures~\cite{silva2019optimization}.

Other researchers focus on generating human-centered explanations which describe the actions of an agent in human-understandable language. 
One such approach is rationale generation which present post-hoc explanations which rationalize the actions taken by an agent in a human-understandable manner~\cite{Ehsan2019AutomatedRG, das2021explainable}. 
Another human-centered explainability methodology is motivated by explaining a model through exposing individual training examples influence the model. 
In instances where data is presented in a format understandable to an end-user, these approaches provide an elegant solution to highlight a meaningful reason for a network's output. 
Prior work has enabled approaches to identify and visualize individually the effect of training examples on the hidden representations of a neural network, and have applied these methods towards explaining the network or understanding the source of bias~\cite{pmlr-v151-silva22a, pmlr-v70-koh17a}. Alternative data-based explainability methods have also provided methods to highlight the sections of the training example which provide a reasoning for an output~\cite{mullenbach-etal-2018-explainable, lakhotia-etal-2021-fid, deyoung-etal-2020-eraser}.

Lastly, an important subfield of explainability which includes both interpretability and human-centered approaches is called Explainable AI Planning (XAIP), which categorize XAI approaches to sequential decision making problems~\cite{ijcai2020p669, hoffmann2019explainable}. 
These approaches present the plan of an algorithm to the explainee. The first set of these approaches seek to reconcile the inference capacity or the mental model~\cite{klein2008macrocognition} of a user, to present model-agnostic explanations to a user. 
Inference reconciliation involves answering investigatory questions from users such as ``Why not action $a$ instead of $a'$?''~\cite{madumal2020explainable, miller_2021}, or ``Why is this plan optimal?''~\cite{khan2009minimal, hayes2017improving}.  
Model reconcilation approaches format explanations to adjust the human's mental model to more accurately align it with the actual conceptual model of the agent~\cite{chakraborti2017plan, sreedharan2019model}. The last category of plan-explanations is policy or behavior summarizations~\cite{amir2019summarizing, lage2019exploring}. These approaches describe the functionality of an AI agent, so that they can be individually applied towards understanding individual actions. 
In this work, we do not present a novel XAI methodology. Within this taxonomy of approaches, our investigation most closely aligns with policy summarizations, i.e. our explanations describe the overall policy of the self-driving car. However, we seek to apply these generated explanations to understand perceptions and usefulness of these policy explanations through a human-subjects study.

\subsection{Evaluating Explainability}
With a greater focus placed on XAI systems, facilitating a means of evaluating the effectiveness and usability of these approaches has become increasingly important.
\textit{Human-grounded evaluation}~\cite{doshi2017towards} is a popular methodology to evaluate the usefulness of proposed approaches within simulated interactions. Human-grounded evaluation seeks to understand the perception of XAI systems and the aspects of the user-experience which can be improved to facilitate smoother interactions with such autonomous agents~\cite{Ehsan2019AutomatedRG, madumal2020explainable, booth2019evaluating, tonekaboni2019clinicians}.
A common practice in human-grounded evaluation is to leverage the principle of mental models \cite{klein2008macrocognition}, wherein researchers attempt to reconcile the differences between the mental model of a user with the conceptual model being explained to measure how well the XAI method explains the agent's model~\cite{hoffman2018metrics, bansal2019updates}.  
This is typically measured by a post-explanation task which attempts to understand how much the explanation has helped the user learn to better understand the AI agent's decisions~\cite{madumal2020explainable, silva2022explainable, zhang2020effect, kenny2021explaining}.
Our user study employs a similar task prediction methodology which reconciles a user's understanding of the self-driving car by asking participants to predict the actions of the car before and after receiving an explanation to measure the affect of an explanation on the accuracy of their predictions. We incorporate confidence ratings to each prediction question to develop a weighted task prediction metric for each participant.

To subjectively evaluate the perception of an XAI methodology, researchers have primarily applied the Technology Acceptance Model (TAM)~\cite{davis1989perceived}. 
Many prior XAI surveys have employed this model to study the willingness of an individual to accept an XAI agent, through metrics such as ease-of-use, usefulness, intention to use, etc. ~\cite{Ehsan2019AutomatedRG,conati2021toward}. 
Another popular avenue of studying acceptance is through the items of trust and satisfaction. In popular prior work, ~\cite{hoffman2018metrics} present a trust scale which predicts whether the XAI system is reliable and believable. 
Recent work also formalizes a new human-AI trust model and emphasizes why ``warranted'' trust is an important factor in XAI acceptance~\cite{jacovi2021formalizing}.
In this paper, we follow these two lines of analysis by leveraging a survey which combines the TAM model for usability with trust to understand the participants' subjective perception of our XAI agents. 

Finally, the last avenue of related work that needs to be covered is studies which pertain to personalization of XAI modules. 
Recent work
highlights the effects of differing XAI modalities on human-AI teaming with respect to subjective and objective metrics~\cite{silva2022explainable}. Their results suggest that explainability alone does not significantly impact trust and compliance; rather adapting to users and ``meeting users half-way'' is a more effective approach for efficient human-AI teaming. 
Prior work has also investigated how factors such as need for cognition~\cite{cacioppo1984efficient}, openness~\cite{goldberg1990alternative} and other personality traits impact design of explainable interfaces for recommender systems~\cite{millecamp2019explain, millecamp2020s}.
Conati et. al. further built on this idea by developing an case study for an XAI interface for intelligent tutoring that personalizes through adaptive hints~\cite{conati2021toward}. 
The need for personalization has been established in prior work, however, many aspects of modeling a user's personal preference towards XAI are still yet to be fully understood. 
There are many other demographic (educational background, learning style, etc.) and situational (domain, stress-levels, etc.) criteria which may still affect personal preference and user-modeling for XAI. 
Through our preference-based XAI user study, we seek to better understand some of these criteria, within the domain of a self-driving car simulation, in order to improve our ability to personalize XAI interfaces. 

\begin{figure}[t]
    \centering
    \includegraphics[width=0.8\textwidth]{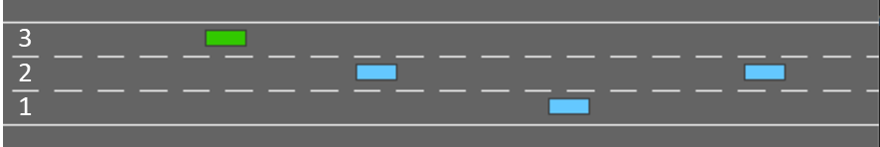}
    \caption{This diagram depicts the environment utilized in this study. The green car denotes the AI agent which is navigating through the highway and presenting explanations to the particpant for each action it takes.}
    \label{fig:highway}
\end{figure}

\section{Methodology}
To analyze utility of different modalities explanations describing the decision making process of a sequential AI-agent, we conducted a novel human-grounded evaluation~\cite{doshi2017towards} experiment to see which explanation modality is the most helpful objectively for simulating/predicting an agent's behavior and subjectively for usability. 
Our study was conducted within the highway domain~\cite{abbeel2004apprenticeship}(see Figure~\ref{fig:highway}). 
In this environment, the car needs to navigate through traffic on a three-lane highway, where the traffic is always moving in the same direction. 
We chose this domain due to the easily understandable nature of the domain. 
Most people would have had prior experience driving or being a passenger in a car on a highway, so they are likely to have an expectation of how to ``properly'' drive on a highway.
This allows us to test whether we are accurately able to convey the car's decision making process, and consolidate the differences between the two. 
Through this study we attempt to not only understand more about explanation preferences and perceptions but also the dispositional (CS Experience, Video Game experience, learning style, etc.) and situational (success/failure) factors which influence these preferences. 
Specifically, our analysis sought to answer the following questions,
\begin{itemize}
{
    \item \textbf{Q1}: Which explanation modality affords the greatest degree of simulatability in terms of understanding the decision making process of the car and accurately predicting the car's actions? 
    \item \textbf{Q2}: How do individual explanation modalities impact subjective measures of usability and trust, and are these individual preferences consistent with the metric of explanation usefulness studied in Q1?
    \item \textbf{Q3}: Are there any interaction affects between dispositional and situational factors, e.g. computer science experience, learning style, success/failure, on the subjective and objective measures studied in this protocol? 
}
\end{itemize}


\begin{table}
\begin{center}
\resizebox{\linewidth}{!}{
\begin{tabular}{ |c|c| } \hline
Category & Questions  \\
\hline
\multirow{3}{10em}{Perceived Usefulness} & Using this explainable agent would be useful for me  \\ 
& Using this explainable agent will improve my effectiveness  \\ 
& Using this explainable agent will improve my performance  \\ 
\hline
\multirow{3}{10em}{Perceived Ease-of-use} & With this explainable agent, it would be easy to get the information I need  \\ 
& Learning to operate with this explainable agent would be easy \\ 
& This explainable agent would be easy to use  \\ 
\hline
\multirow{4}{10em}{Attitude} & Using this explainable agent is an idea I like  \\ 
& Using this explainable agent would be a pleasant experience \\ 
& Using this explainable agent is a good idea  \\ 
& Using this explainable agent is a wise idea  \\ 
\hline
\multirow{3}{10em}{Trust} & I trust this explainable agent  \\ 
& This explainable agent is reliable  \\ 
& This explainable agent is trustworthy  \\ 
\hline
\multirow{3}{10em}{Intention to use} & When I will need it, I will intend to use this explainable agent rather than an agent with no explanation  \\ 
& When I will need it, I predict I would use this explainable agent rather than an agent with no explanation  \\ 
& When I will need it, I would like to use this explainable agent rather than an agent with no explanation \\ 
\hline
\end{tabular}
}
\end{center}
\caption{\label{tab:survey}This table depicts the specific questions asked within each item of the usability and trust questionnaire employed in this study. This survey was previously used to measure perception of ``e-services.'' In our study, we replaced all references of ``e-services'' to ``explainable agent.''}
\end{table}


\subsection{Experiment Design}
The factors in our experiment were (1) Explanation Format and (2) Success-vs.-Failure video. Our experiment was a between-subjects study with a 2$\times$4 study design. 

\textbf{Explanation Format} --
We chose XAI modalities that can all be generated from a single methodology presented in prior work~\cite{silva2020neuralencoding}. This method converts learned policies into discretized decision trees which elucidate the decision making process within an AI-agent's policy~\cite{silva2019optimization}. 
As such, we chose four explanation modalities that can be generated interchangeably through this base model. These modalities are as follows 
\begin{enumerate}
    \item \textbf{Tree}: A decision tree which encodes the self-driving car's policy.
    \item \textbf{Basic Text}: A policy description generated using a template from the decision-tree policy.
    \item \textbf{Modified Text}: A modified version of the text description, presented in a format which is easier to parse, with simplified language and indentation. 
    \item \textbf{Program}: A set of \emph{if-else} statements encoding the logic of the decision tree. 
\end{enumerate}
The selection of these modalities was motivated by recent explainable AI work for explaining the policies of sequential decision making systems. Decision trees have become a popular method of explaining decisions for human-AI teaming scenarios~\cite{paleja2021utility}. Differential decision trees have been proven to be an interpretable method of representing a policies that can be employed towards generating ``white-box'' explanations for users that actually represent the underlying behavior~\cite{silva2020neuralencoding}. 
Language has also shown to be an effective means of  explaining the actions of an sequential decision making agent~\cite{hayes2017improving, das2021explainable, Ehsan2019AutomatedRG}. The motivation for including two versions of a language baseline is that the ``Modified Text'' baseline serves as an ``Abstraction''~\cite{ijcai2020p669} for the raw text explanation, i.e. it formats the explanation in a more human-like manner such that it is more palatable to the user.
Finally, the choice of program/rule-based explanations was to cater to scenarios wherein explainable systems are utilized to assist domain experts who wish to debug agent behavior. We were interested in understanding if computer science experience played a role in determining whether participants were able to process the same information better as psuedo-code or in a language format. 
The specific explanations provided to participants in our study are provided in the Appendix (Section: ~\ref{sec:Explanation_examples})

\textbf{Success/Failure} --
Prior work has studied how the nature of an explanation sways a user's ability to tolerate the agent failing~\cite{Ehsan2019AutomatedRG}. In this study, we analyze the opposite relationship, i.e. how does seeing the agent succeed or fail impact their perception of the explanation. At the start of the experiment, each participant was shown a one-minute video of a simulation of the car in the highway domain to help the participant build a mental model the AI's behavior. Each participant was shown one of two videos. In the first video, the car crashed at the end of the video (failure), and, in the second, the car reaches the ``finish line'' (success). 
Both these videos were generated using the same policy for the agent. 
Our aim was to measure whether watching the car succeed or crash subjectively influenced participants' perception of any XAI modality or objectively impaired their ability to apply the explanation.  


\subsection{Metrics}
In this section we describe the metrics we employ to subjectively gauge perception of each explanation, with regards to usability and trust, and objectively evaluate a user's ability to simulate the decision making of the car. 
To measure usability, we adapted a survey, which incorporated trust into the TAM model~\cite{davis1989perceived}, from prior work on human-evaluation of e-service systems~\cite{belanche2012integrating}. This survey included questions on \textit{usability}, \textit{ease of use}, \textit{attitude}, \textit{intention to use}, and \textit{trust}.
We replaced references to ``e-service'' in the original survey with ``explainable agent'' for this user study. The complete survey utilized in this study can be viewed in Table~\ref{tab:survey}.
Note that we did not employ the frequently utilized trust scale for XAI proposed in ~\cite{hoffman2018metrics}, because our study did not satisfy the assumptions required as per the authors, i.e. ``the participant has had considerable experience using the XAI system.''

To measure objective simulatability, we computed a prediction score using participants' predictions before and after receiving an explanation for the car's actions. We asked participants four prediction questions where they predicted the next sequence of actions the car will take. Using their answers to these questions, we compute a task prediction score as shown in Equation~\ref{eq:score},
\begin{equation}
\label{eq:score}
score = \sum_{i=1}^{4} c_{a,i} \times \delta_{a,i} - \sum_{i=1}^{4} c_{b,i} \times \delta_{b,i}
\end{equation}
In this formula, the $\delta$ parameters represent whether or not the participant was able to predict the car's actions correctly.
$\delta_{a,i}$ is assigned a value of $+1$ if the $i^{th}$ question was answered correctly prior to receiving an explanation and $-1$ otherwise. $\delta_{b,i}$ similarly represents the correctness of the participant's prediction for the $i^{th}$ question after receiving an explanation.
$c_{b,i}$ represents the confidence rating for question, $i$, before receiving an explanation, and $c_{a,i}$ represents the confidence rating for the $i^{th}$ question after receiving an explanation. 
Confidence ratings are obtained by asking participants how confident they are in their prediction of the car's next sequence of actions, on a 5-item scale from ``Not confident at all'' to ``Extremely confident.''
To compute $c_{b,i}$ and $c_{a,i}$, we assign numeric values to a participants confidence rating uniforming between 0.2 and 1, in increments of 0.2 (i.e., Not confident $=$ 0.2, Slightly confident $=$ 0.4, Moderately confident $=$ 0.6, Very confident $=$ 0.8, Extremely confident $=$ 1). 
When combined, $c$ and $\delta$ represent a weighted prediction score. The score variable represents the difference between the weighted prediction scores across four different prediction questions. 
We also measure the unweighted score, the number of correct answers in phase 2, and weighted number of correct answers in phase 2 to track simulatability performance.

\subsection{Procedure}
This experiment was conducted online via Amazon Mechanical Turk. Our study began with a demographics survey about age, gender, education and experience with computer science and video games.
Computer Science and video game experience were measured on a 4-point, self-reported scale from ``very inexperienced'' to ``very experienced.''
Participants were also asked to answer short surveys regarding their orientation towards things or people~\cite{graziano2012orientations} and learning style (visual-vs.-verbal~\cite{mayer2003three}).  
The rest of our study is divided into three phases. 

In Phase 1 of the study, the participant first received a one-minute video of the car driving on a highway, in which the car would either reach the finish line (success) or crash into another car (failure) at the end of the video, in order to prime the participant with the car's behavior.
Next, the participant would be asked to complete four prediction tasks. 
For each prediction task, the participant was shown a unique, eight-second video of the car driving (i.e. overtaking/changing lanes/slowing down, etc.) on the virtual highway.  
Based on this video of the car, participants were asked to predict the next set of actions of the agent from a set of five options (including an option for none of the above), by utilizing their inferred mental model of the car. 
Each scenario involved the car executing a policy on a different part of the highway.
Each prediction question was accompanied with a 5-point confidence rating (Not confident - Extremely confident). 

In Phase 2, participants would perform the same tasks as in Phase 1, with the exception being that participants also received an additional explanation for the actions of the car in one of the four formats specified earlier. Conducting the same prediction tasks with and without an explanation allowed us to directly analyze the impact of an explanation on the perception of the explainable agent. 
Finally, in the third and final phase, participants were asked to complete our usability and trust survey to subjectively evaluate the explanation modality they worked with.   
We adapted a survey which incorporated trust into the TAM model~\cite{davis1989perceived}, from prior work on human-evaluation of e-service systems~\cite{belanche2012integrating} involving questions on \textit{usability}, \textit{ease-of-use}, \textit{attitude}, \textit{intention to use}, and \textit{trust}. 
Note that we did not employ the frequently utilized trust scale for XAI proposed by ~\cite{hoffman2018metrics}, because out study did not satisfy the assumptions required as per the authors, i.e. ``the participant has had considerable experience using the XAI system.''
In our case, participants were interacting with the XAI agent for the first time for only 20 minutes, therefore we ascertained that this scale was not applicable. 
Our study design relates to that of another from prior work~\cite{booth2019evaluating}, in the context of understanding how to communicate policies, of a sequential decision making agent, to end-users. 
The authors of that prior work~\cite{booth2019evaluating} sought to understand the impact different representation languages when used to encode policies, where each explanation to a policy was presented in the form of a representation language.
In contrast, we seek to understand how varying the modality of the explanation impacts perceived interpretability and usability of the policy explanation, which is crucial for facilitating large-scale adoption of XAI systems. 



\begin{figure}[!ht]
  \centering
  \subfloat[][]{\includegraphics[width=.31\textwidth]{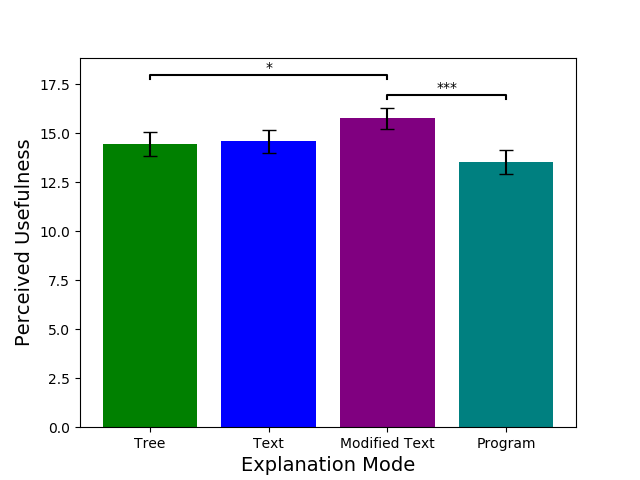}}\quad
  \subfloat[][]{\includegraphics[width=.31\textwidth]{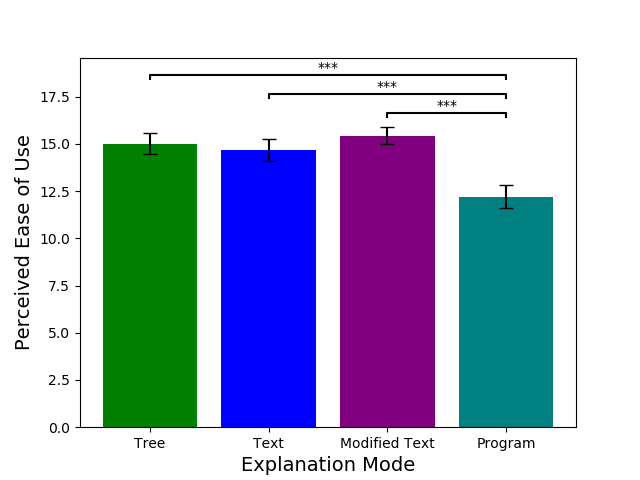}}\\
  \subfloat[][]{\includegraphics[width=.31\textwidth]{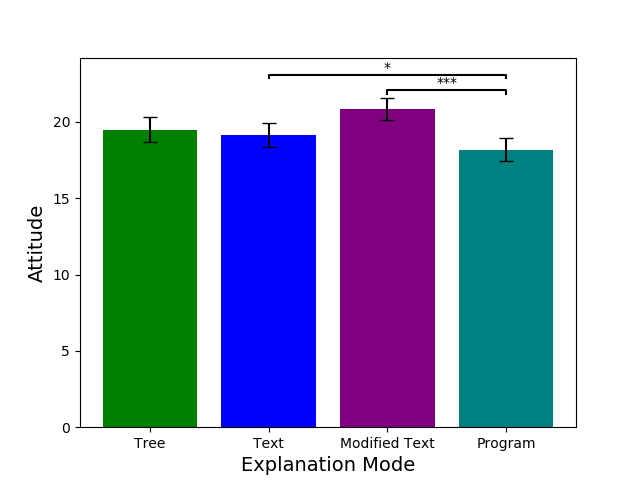}}\quad
  \subfloat[][]{\includegraphics[width=.31\textwidth]{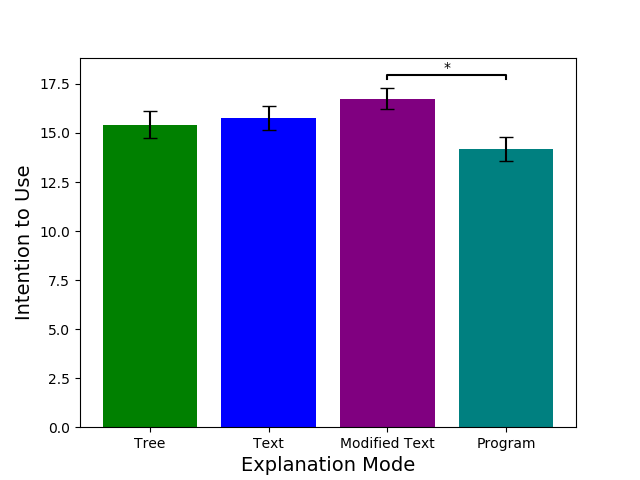}}\quad
  \subfloat[][]{\includegraphics[width=.31\textwidth]{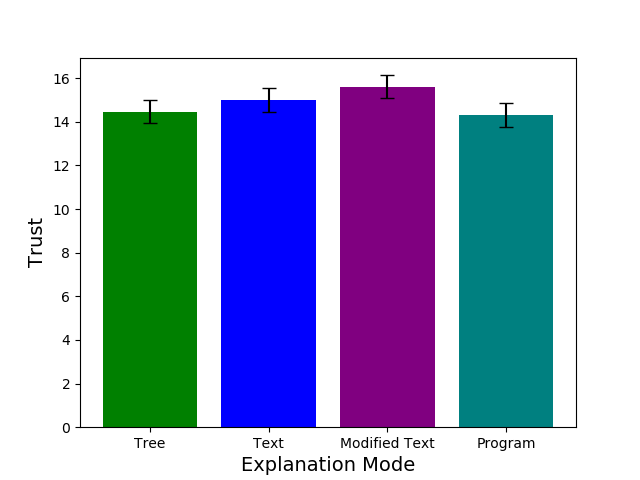}}
  \caption{  These graphs plot the means and standard errors for each subjective evaluation metric across the four explanation modalities. Significant differences between modalities is noted in the graphs. }
  \label{fig:sub1}
\end{figure}

\subsection{Results}

Our analysis was conducted on data from 231 participants, recruited from mechanical turk (54\% identified as Male, 46\% identified as Female, and $<$1\% identified as Non-binary/other).
Out of all responses collected, we only included responses in our final dataset from participants who had submitted the survey once. 
To the best of our knowledge, all responses that were from repeat or malicious responders were filtered out. 
A total of 46 participants reported having some degree of computer science experience. 
The average time taken for our survey was 18 minutes and participants were paid \$4 for completing our study (which equates to \$13.34 per hour).

We created a multivariate regression model with the explanation mode, success/failure and demographics values as the independent variable, with the dependent variable being the subjective or objective metric being studied. 
Models were checked to meet normality and homoscedasticity assumptions. Omnibus tests were performed before pairwise comparisons were made. We used multivariate linear regression with AIC as our occam's razor for modelling covariates and interaction effects.
For Q1 (understanding the simulatability of each individual modality), we compared how each explanation mode affected the task prediction performance using our objective metrics. We find that tree explanations were significantly more beneficial for predicting more questions correctly in phase 2 when compared to modified (Estimate = -1.257, SE = 0.374, p $<$ 0.001) and basic text (Estimate = -1.138, Standard Error (SE) = 0.3548, p $<$ 0.01). 
After taking into account confidence ratings, the usage of tree explanations still significantly improved weighted number of correct answers in Phase 2 as compared to the modified text explanations (Estimate = -0.571, SE = 0.167, p $<$ 0.001), and the basic text-based explanations (Estimate = -0.480, SE = 0.160, p $<$ 0.01). 
For the score metric described earlier, both trees (Estimate = 2.517, SE = 0.558, p $<$ 0.001) and programs (Estimate = 1.414, SE = 0.555, p $<$ 0.05) significantly improved the participant's score when compared to the modified text baseline. \textbf{These results imply that users are able to more accurately simulate and understand an agent's decisions using trees.} 
 

With respect to Q2 (understanding the perceived usability of individual modalities), we found that modified text was rated to be significantly more useful than both the program (Estimate = 47.6434, SE = 12.484, p $<$ 0.001) and the tree (Estimate = 29.098, SE = 12.470, p $<$ 0.05) baselines. 
For ease of use, the tree, text, and modified text baselines were rated significantly higher than the program explanation (p $<$ 0.001).
For the metric of intention to use, a Wilcoxon signed rank test showed that modified text was preferred to program (p $<$ 0.05). 
These results suggest an inconsistency between the subjective and objective evaluation metrics for the decision tree and program vs text-based modalities. 
\textbf{Although they were found to be less useful for accurately predicing the actions of the car, participants perceived text-based explanations as significantly more usable than decision trees and programs.}
An important point to note with respect to Q2 is that the model for trust did not pass the assumptions for our parametric test. 
Consequently, we conducted a non-parametric analysis for trust, however, none of the explanation modalities were found to significantly impact trust. 
In relation to Q3 (understanding the effect of individual factors on the subjective and objective measures of each modality), 
we found that participants with low CS experience have significantly improved relative prediction scores when using the modified text explanation as compared to the tree (Estimate = -2.1597, SE = 0.611, p $<$ 0.001) or program (Estimate = -1.436, SE = 0.609, p $<$ 0.05) explanations.
For usefulness, higher CS experience significantly decreases the relative advantage of text over program for both the basic (Estimate = -14.12, SE = 6.335, p $<$ 0.05) and modified text (Estimate = -19.69, SE = 6.597, p $<$ 0.01) modalities. 
Similarly, with respect to attitude and ease-of-use, high-CS experience was found to significantly decrease the preference of modified text (p < 0.01) and text (p < 0.05) explanations relative to programs. 
\textbf{Overall, our results showed that with respect to simulatability and usability, increasing CS experience negatively impacts the text-based explanations compared to the program or tree explanations.} 

We note that we did not find self-reported learning-style preference (visual vs verbal) to be a significant influencing factor for either the subjective or objective measures we studied.  
Next we studied whether success and failure were found to significantly influence XAI perception. 
With respect to success, we found that watching the car succeed in the priming video -- as opposed to failing, i.e. crashing --  significantly improved a participant's attitude (Estimate = 11.986, SE = 4.64, p $<$ 0.05). However, this effect was not observed for the other dependent variables, i.e. ease-of-use, trust, usefulness and intent-to-use. This implies that although, on average, participants felt that working with the XAI agent that failed was ``unpleasant'', it did not impact their usability. This may indicate that better care needs to taken to appease end-users in situations where they work with agents that frequently fail. Unlike in the case of attitude, success/failure was not found to affect the score of a participant, i.e. watching the car fail did not affect the participants ability to understand the explanation. 



\subsection{Qualitative insights and Discussion}

At the end of Phase 2, we asked participants to describe their experience working with the format of explanation they received in order to gain some insight into the their preferences, via an open-ended textual prompt. One important finding was that participants often reacted adversely to receiving an explanation the participant did not understand or could not parse. This trend was particularly prevalent for the program-based explanation. 
People with little experience of programming were often discouraged and confused by the program-based explanation. One participant stated that the explanation was counter-productive in that it made the participant ``second guess [their] initial choice,'' and further stated that ``If it was supposed to be reassuring and confirming, it wasn't.'' Another participant stated that the nature of the program-based explanation made it ``functionally useless'' to the task assigned. 

Yet, other participants seemed to value conciseness within the explanations over the detailed nature of the text-based explanations. For example, one participant took issue with the modified-text explanation as it was not ordered in a way that was easy to understand. The participant stated, ``Some of the sentences could have been combined and just said left or right instead of having a statement for each.'' Another participant stated that they were ``better with visual learning,'' and, therefore, preferred to go by their initial assumptions based on the video rather than use the detailed text explanation of the car's policy indicating that ill-fitting explanations made the participant choose to ignore the explanation provided. \color{black}
A third participant expressed their struggles in learning to use the decision tree explanation and suggested that, "it would have been helpful if there was animated instructions on the original video to show the rules."
Thereby indicating that the participant would be more amenable to using this format of explanation if it was presented in a way that was easier to parse.   


\textbf{These examples highlight that when explanations are ill-fitting of an individual's dispositional or situational circumstances, users may be unable or unwilling to utilize the explanations to adapt their understanding of the car.}
Humans create mental models for systems they interact with~\cite{hoffman2018metrics}, and an accessible explanation should focus on identifying and correcting the misconceptions within this mental model in a way which caters to the socio-technical disposition of the user. One participant's response encapsulates this sentiment: After receiving the decision tree the participant stated, the explanation ``was generally helpful in that it helped [the participant] focus on the other car that was the biggest factor in the AI's decision making.'' This indicated that the participant was able to apply the explanation to improve their mental model of the car's behavior, by identifying the factors in the environment that influence the car's decisions. Another participant stated that the modified-text explanation ``really helped'' because ``it showed how to see the car and how it would interact with the world around it.'' In both these situations, the user was more open to adopting the explanation because the explanation was able to satisfactorily fill in the gaps in their mental model of the car, by helping participants perceive how the car may be processing the information available in the environment to make decisions. 

\textbf{In our study, we found inconsistency in human preferences of explanation modalities with respect to subjective and objective metrics.}
Participants found language-based explanations to be significantly more useful (p $<$ 0.001) even though  participants performed better according to our objective metrics when using the tree-based explanation (p $<$ 0.001). 
This contradiction implies that being able to successfully apply an explanation does not necessarily enhance a user's assessment of the XAI system. 
These findings support discourse from prior work on human-centered or user-centered perspectives to explainability~\cite{liao2020questioning, ehsan2020human, dhanorkar2021needs}. 
Participants' preference towards using modes of explanation which objectively perform poorer on task performance metrics is a clear indicator that explanations need to consider the individual dispositions of the potential end-user to engender adoption. 
Our results support the position that researchers should design
personalized XAI interfaces which can cater to the social needs of the end-users interacting with these systems. 
We do not claim to be the first to show that Personalized XAI is necessary, which has already been shown in prior work~\cite{millecamp2019explain, millecamp2020s}. 
However, these works are restricted to recommendation/tutoring systems and were conducted on a much smaller sample population (the other two studies were conducted on 30/72 participants whereas this study was conducted on 231 participants).
We contribute to the existing body of literature for a novel domain, by providing insights for potentially developing XAI interfaces for a sequential decision making task like a self-driving car simulator. 
Our analysis identifies key demographics factors such as computer science experience, and highlighted the importance of these factors for with respect to subjective perception and objective use of XAI modalities. 

\section{Limitations and Future Work}

Firstly, our study follows a human-grounded evaluation structure we performed our analysis on for a simulated self-driving car. Therefore, it is important to acknowledge that while these results provide a comprehensive initial estimate, they may vary when this study is replicated on the real task. Future work could benefit from a similar study in a real setting, i.e. application grounded evaluation, which accurately reproduces the experience of receiving explanations from a self-driving car. Secondly, this study does not consider longitudinal human-adaptation to XAI systems. It may be possible that although a participant initially does not prefer to use an explanation after brief interactions with the agent, a period of time to adapt to the explanation modality may alter their preference. In future work, we aim to study how personalized XAI perceptions are influenced by longitudinal factors and prolonged interactions with the XAI agent. 
Another relevant topic to discuss with regards to adaption is the cost of adaption. It may be worthwhile to study whether a user prefers to be taught how to adapt to a ``useful'' explanation or whether the explanation should be modified to cater to them. Intuitively, one would expect that users would prefer having a personalized explanation for them, but future work could empirically study this trade-off.

Another important limitation to consider is the participant's level of immersion. Viewing the simulation of the car in our environment may not be enough for the participant to recognize the consequences of working with a self-driving car. In a situation where the stakes are more obvious, participants may perceive explanations differently.
This may have contributed towards the lack of significance for the trust model. 
Since participants may not have been immersed/understood the potential real-world consequences, their internal model for trust may have been independent to the explanation provided. 
However, we believe that our study still contributes novel insights that provide a stepping stone towards a future application-grounded analysis.  Finally, in future work, we aim to leverage the insights from this study to develop a personalizable XAI methodology. 
In such a case, the XAI interface could evaluate an individual's disposition and demographic factors to recommend a type of explanation.
Then the user can specify any additional properties they would like within the explanation and adaptively modify the type of explanations it receives from the agent. 
We hypothesize that such an approach would truly give rise to human-centered explainability and bridge the gap between stakeholders and AI technology. 
\section{Conclusion}
Explainable AI must have a stakeholder-focus to engender long-term adoption. 
Simply unraveling the internal mechanisms of an Artificial Intelligence agent is insufficient if it is not presented in a way the end-user can easily understand. 
To produce user-centered XAI approaches, we need to better understand what influences XAI perception. 
In this paper, we present a novel user-study which studies subjective user-preference towards disparate XAI modalities, for a self-driving car, and how situational (e.g., watching the car succeed or fail) and dispositional factors (e.g., computer science experience) influence this perception. 
We show that computer science experience can reduce an individual's perception towards the text-based modalities, as well as how watching the car fail (crash into another car) worsens their attitude towards the XAI agent.
Our findings also highlight an important internal inconsistency in explanation preference. Text-based explanations were perceived to be more usable according to our subjective survey, however, decision tree explanations were found to be more useful in terms of more accurately predicting the car's actions. 
XAI developers need to balance the tradeoff between willingness to adopt and usefulness, as the perceived usability varies based on an individuals specific intrinsic and situational criteria.
We hope that this work promotes a wider study of personalized XAI approaches which curate explanations to fit the particular needs and circumstances of individual stakeholders. 

\bibliographystyle{unsrt}  
\bibliography{references}  

\begin{thebibliography}{10}

\bibitem{matthews2019individual}
Gerald Matthews, Jinchao Lin, April~Rose Panganiban, and Michael~D Long.
\newblock Individual differences in trust in autonomous robots: Implications
  for transparency.
\newblock {\em IEEE Transactions on Human-Machine Systems}, 50(3):234--244,
  2019.

\bibitem{das2021explainable}
Devleena Das, Siddhartha Banerjee, and Sonia Chernova.
\newblock Explainable ai for robot failures: Generating explanations that
  improve user assistance in fault recovery.
\newblock In {\em Proceedings of the 2021 ACM/IEEE International Conference on
  Human-Robot Interaction}, pages 351--360, 2021.

\bibitem{kwon2018expressing}
Minae Kwon, Sandy~H Huang, and Anca~D Dragan.
\newblock Expressing robot incapability.
\newblock In {\em Proceedings of the 2018 ACM/IEEE International Conference on
  Human-Robot Interaction}, pages 87--95, 2018.

\bibitem{robinette2017effect}
Paul Robinette, Ayanna~M Howard, and Alan~R Wagner.
\newblock Effect of robot performance on human--robot trust in time-critical
  situations.
\newblock {\em IEEE Transactions on Human-Machine Systems}, 47(4):425--436,
  2017.

\bibitem{xu2019explainable}
Feiyu Xu, Hans Uszkoreit, Yangzhou Du, Wei Fan, Dongyan Zhao, and Jun Zhu.
\newblock Explainable ai: A brief survey on history, research areas, approaches
  and challenges.
\newblock In {\em CCF international conference on natural language processing
  and Chinese computing}, pages 563--574. Springer, 2019.

\bibitem{jacovi2021formalizing}
Alon Jacovi, Ana Marasovi{\'c}, Tim Miller, and Yoav Goldberg.
\newblock Formalizing trust in artificial intelligence: Prerequisites, causes
  and goals of human trust in ai.
\newblock In {\em Proceedings of the 2021 ACM conference on fairness,
  accountability, and transparency}, pages 624--635, 2021.

\bibitem{grath2018interpretable}
Rory~Mc Grath, Luca Costabello, Chan~Le Van, Paul Sweeney, Farbod Kamiab, Zhao
  Shen, and Freddy Lecue.
\newblock Interpretable credit application predictions with counterfactual
  explanations.
\newblock {\em arXiv preprint arXiv:1811.05245}, 2018.

\bibitem{pawar2020explainable}
Urja Pawar, Donna O’Shea, Susan Rea, and Ruairi O’Reilly.
\newblock Explainable ai in healthcare.
\newblock In {\em 2020 International Conference on Cyber Situational Awareness,
  Data Analytics and Assessment (CyberSA)}, pages 1--2. IEEE, 2020.

\bibitem{anjomshoae2019explainable}
Sule Anjomshoae, Amro Najjar, Davide Calvaresi, and Kary Fr{\"a}mling.
\newblock Explainable agents and robots: Results from a systematic literature
  review.
\newblock In {\em 18th International Conference on Autonomous Agents and
  Multiagent Systems (AAMAS 2019), Montreal, Canada, May 13--17, 2019}, pages
  1078--1088. International Foundation for Autonomous Agents and Multiagent
  Systems, 2019.

\bibitem{samek2017explainable}
Wojciech Samek, Thomas Wiegand, and Klaus-Robert M{\"u}ller.
\newblock Explainable artificial intelligence: Understanding, visualizing and
  interpreting deep learning models.
\newblock {\em arXiv preprint arXiv:1708.08296}, 2017.

\bibitem{simonyan2013deep}
Karen Simonyan, Andrea Vedaldi, and Andrew Zisserman.
\newblock Deep inside convolutional networks: Visualising image classification
  models and saliency maps.
\newblock {\em arXiv preprint arXiv:1312.6034}, 2013.

\bibitem{humbird2018deep}
Kelli~D Humbird, J~Luc Peterson, and Ryan~G McClarren.
\newblock Deep neural network initialization with decision trees.
\newblock {\em IEEE transactions on neural networks and learning systems},
  30(5):1286--1295, 2018.

\bibitem{silva2020neuralencoding}
Andrew Silva and Matthew Gombolay.
\newblock Neural-encoding human experts' domain knowledge to warm start
  reinforcement learning, 2020.

\bibitem{letham2015interpretable}
Benjamin Letham, Cynthia Rudin, Tyler~H McCormick, and David Madigan.
\newblock Interpretable classifiers using rules and bayesian analysis: Building
  a better stroke prediction model.
\newblock {\em The Annals of Applied Statistics}, 9(3):1350--1371, 2015.

\bibitem{ghaeini2018interpreting}
Reza Ghaeini, Xiaoli~Z Fern, and Prasad Tadepalli.
\newblock Interpreting recurrent and attention-based neural models: a case
  study on natural language inference.
\newblock {\em arXiv preprint arXiv:1808.03894}, 2018.

\bibitem{amir2019summarizing}
Ofra Amir, Finale Doshi-Velez, and David Sarne.
\newblock Summarizing agent strategies.
\newblock {\em Autonomous Agents and Multi-Agent Systems}, 33(5):628--644,
  2019.

\bibitem{lage2019exploring}
Isaac Lage, Daphna Lifschitz, Finale Doshi-Velez, and Ofra Amir.
\newblock Exploring computational user models for agent policy summarization.
\newblock In {\em IJCAI: proceedings of the conference}, volume~28, page 1401.
  NIH Public Access, 2019.

\bibitem{chakraborti2017balancing}
Tathagata Chakraborti, Sarath Sreedharan, and Subbarao Kambhampati.
\newblock Balancing explicability and explanation in human-aware planning.
\newblock {\em arXiv preprint arXiv:1708.00543}, 2017.

\bibitem{wang2019designing}
Danding Wang, Qian Yang, Ashraf Abdul, and Brian~Y Lim.
\newblock Designing theory-driven user-centric explainable ai.
\newblock In {\em Proceedings of the 2019 CHI conference on human factors in
  computing systems}, pages 1--15, 2019.

\bibitem{ehsan2020human}
Upol Ehsan and Mark~O Riedl.
\newblock Human-centered explainable ai: Towards a reflective sociotechnical
  approach.
\newblock {\em arXiv preprint arXiv:2002.01092}, 2020.

\bibitem{ehsan2021expanding}
Upol Ehsan, Q~Vera Liao, Michael Muller, Mark~O Riedl, and Justin~D Weisz.
\newblock Expanding explainability: Towards social transparency in ai systems.
\newblock In {\em Proceedings of the 2021 CHI Conference on Human Factors in
  Computing Systems}, pages 1--19, 2021.

\bibitem{zhou2022exsum}
Yilun Zhou, Marco~Tulio Ribeiro, and Julie Shah.
\newblock Exsum: From local explanations to model understanding.
\newblock {\em arXiv preprint arXiv:2205.00130}, 2022.

\bibitem{ghassemi2021false}
Marzyeh Ghassemi, Luke Oakden-Rayner, and Andrew~L Beam.
\newblock The false hope of current approaches to explainable artificial
  intelligence in health care.
\newblock {\em The Lancet Digital Health}, 3(11):e745--e750, 2021.

\bibitem{liao2020questioning}
Q~Vera Liao, Daniel Gruen, and Sarah Miller.
\newblock Questioning the ai: informing design practices for explainable ai
  user experiences.
\newblock In {\em Proceedings of the 2020 CHI Conference on Human Factors in
  Computing Systems}, pages 1--15, 2020.

\bibitem{conati2021toward}
Cristina Conati, Oswald Barral, Vanessa Putnam, and Lea Rieger.
\newblock Toward personalized xai: A case study in intelligent tutoring
  systems.
\newblock {\em Artificial Intelligence}, 298:103503, 2021.

\bibitem{millecamp2020s}
Martijn Millecamp, Nyi~Nyi Htun, Cristina Conati, and Katrien Verbert.
\newblock What's in a user? towards personalising transparency for music
  recommender interfaces.
\newblock In {\em Proceedings of the 28th ACM Conference on User Modeling,
  Adaptation and Personalization}, pages 173--182, 2020.

\bibitem{silva2022explainable}
Andrew Silva, Mariah Schrum, Erin Hedlund-Botti, Nakul Gopalan, and Matthew
  Gombolay.
\newblock Explainable artificial intelligence: Evaluating the objective and
  subjective impacts of xai on human-agent interaction.
\newblock {\em International Journal of Human--Computer Interaction}, pages
  1--15, 2022.

\bibitem{davis1989perceived}
Fred~D Davis.
\newblock Perceived usefulness, perceived ease of use, and user acceptance of
  information technology.
\newblock {\em MIS quarterly}, pages 319--340, 1989.

\bibitem{stilgoe2019self}
Jack Stilgoe.
\newblock Self-driving cars will take a while to get right.
\newblock {\em Nature Machine Intelligence}, 1(5):202--203, 2019.

\bibitem{zablocki2021explainability}
{\'E}loi Zablocki, H{\'e}di Ben-Younes, Patrick P{\'e}rez, and Matthieu Cord.
\newblock Explainability of vision-based autonomous driving systems: Review and
  challenges.
\newblock {\em arXiv preprint arXiv:2101.05307}, 2021.

\bibitem{yosinski2015understanding}
Jason Yosinski, Jeff Clune, Anh Nguyen, Thomas Fuchs, and Hod Lipson.
\newblock Understanding neural networks through deep visualization.
\newblock {\em arXiv preprint arXiv:1506.06579}, 2015.

\bibitem{selvaraju2017grad}
Ramprasaath~R Selvaraju, Michael Cogswell, Abhishek Das, Ramakrishna Vedantam,
  Devi Parikh, and Dhruv Batra.
\newblock Grad-cam: Visual explanations from deep networks via gradient-based
  localization.
\newblock In {\em Proceedings of the IEEE International Conference on Computer
  Vision}, pages 618--626, 2017.

\bibitem{adebayo2018sanity}
Julius Adebayo, Justin Gilmer, Michael Muelly, Ian Goodfellow, Moritz Hardt,
  and Been Kim.
\newblock Sanity checks for saliency maps.
\newblock {\em Advances in neural information processing systems}, 31, 2018.

\bibitem{serrano2019attention}
Sofia Serrano and Noah~A Smith.
\newblock Is attention interpretable?
\newblock {\em arXiv preprint arXiv:1906.03731}, 2019.

\bibitem{kindermans2019reliability}
Pieter-Jan Kindermans, Sara Hooker, Julius Adebayo, Maximilian Alber, Kristof~T
  Sch{\"u}tt, Sven D{\"a}hne, Dumitru Erhan, and Been Kim.
\newblock The (un) reliability of saliency methods.
\newblock In {\em Explainable AI: Interpreting, Explaining and Visualizing Deep
  Learning}, pages 267--280. Springer, 2019.

\bibitem{paleja2021utility}
Rohan Paleja, Muyleng Ghuy, Nadun Ranawaka~Arachchige, Reed Jensen, and Matthew
  Gombolay.
\newblock The utility of explainable ai in ad hoc human-machine teaming.
\newblock {\em Advances in Neural Information Processing Systems}, 34:610--623,
  2021.

\bibitem{silva2019optimization}
Andrew Silva, Matthew Gombolay, Taylor Killian, Ivan Jimenez, and Sung-Hyun
  Son.
\newblock Optimization methods for interpretable differentiable decision trees
  applied to reinforcement learning.
\newblock volume 108 of {\em Proceedings of Machine Learning Research}, pages
  1855--1865, Online, 26--28 Aug 2020. PMLR.

\bibitem{Ehsan2019AutomatedRG}
Upol Ehsan, Pradyumna Tambwekar, Larry Chan, Brent Harrison, and Mark~O. Riedl.
\newblock Automated rationale generation: a technique for explainable ai and
  its effects on human perceptions.
\newblock In {\em IUI '19}, 2019.

\bibitem{pmlr-v151-silva22a}
Andrew Silva, Rohit Chopra, and Matthew Gombolay.
\newblock Cross-loss influence functions to explain deep network
  representations.
\newblock In Gustau Camps-Valls, Francisco J.~R. Ruiz, and Isabel Valera,
  editors, {\em Proceedings of The 25th International Conference on Artificial
  Intelligence and Statistics}, volume 151 of {\em Proceedings of Machine
  Learning Research}, pages 1--17. PMLR, 28--30 Mar 2022.

\bibitem{pmlr-v70-koh17a}
Pang~Wei Koh and Percy Liang.
\newblock Understanding black-box predictions via influence functions.
\newblock In Doina Precup and Yee~Whye Teh, editors, {\em Proceedings of the
  34th International Conference on Machine Learning}, volume~70 of {\em
  Proceedings of Machine Learning Research}, pages 1885--1894. PMLR, 06--11 Aug
  2017.

\bibitem{mullenbach-etal-2018-explainable}
James Mullenbach, Sarah Wiegreffe, Jon Duke, Jimeng Sun, and Jacob Eisenstein.
\newblock Explainable prediction of medical codes from clinical text.
\newblock In {\em Proceedings of the 2018 Conference of the North {A}merican
  Chapter of the Association for Computational Linguistics: Human Language
  Technologies, Volume 1 (Long Papers)}, pages 1101--1111, New Orleans,
  Louisiana, June 2018. Association for Computational Linguistics.

\bibitem{lakhotia-etal-2021-fid}
Kushal Lakhotia, Bhargavi Paranjape, Asish Ghoshal, Scott Yih, Yashar Mehdad,
  and Srini Iyer.
\newblock {F}i{D}-ex: Improving sequence-to-sequence models for extractive
  rationale generation.
\newblock In {\em Proceedings of the 2021 Conference on Empirical Methods in
  Natural Language Processing}, pages 3712--3727, Online and Punta Cana,
  Dominican Republic, November 2021. Association for Computational Linguistics.

\bibitem{deyoung-etal-2020-eraser}
Jay DeYoung, Sarthak Jain, Nazneen~Fatema Rajani, Eric Lehman, Caiming Xiong,
  Richard Socher, and Byron~C. Wallace.
\newblock {ERASER}: {A} benchmark to evaluate rationalized {NLP} models.
\newblock In {\em Proceedings of the 58th Annual Meeting of the Association for
  Computational Linguistics}, pages 4443--4458, Online, July 2020. Association
  for Computational Linguistics.

\bibitem{ijcai2020p669}
Tathagata Chakraborti, Sarath Sreedharan, and Subbarao Kambhampati.
\newblock The emerging landscape of explainable automated planning \& decision
  making.
\newblock In Christian Bessiere, editor, {\em Proceedings of the Twenty-Ninth
  International Joint Conference on Artificial Intelligence, {IJCAI-20}}, pages
  4803--4811. International Joint Conferences on Artificial Intelligence
  Organization, 7 2020.
\newblock Survey track.

\bibitem{hoffmann2019explainable}
J{\"o}rg Hoffmann and Daniele Magazzeni.
\newblock Explainable ai planning (xaip): overview and the case of contrastive
  explanation.
\newblock {\em Reasoning Web. Explainable Artificial Intelligence}, pages
  277--282, 2019.

\bibitem{klein2008macrocognition}
Gary Klein and Robert~R Hoffman.
\newblock Macrocognition, mental models, and cognitive task analysis
  methodology.
\newblock {\em Naturalistic decision making and macrocognition}, pages 57--80,
  2008.

\bibitem{madumal2020explainable}
Prashan Madumal, Tim Miller, Liz Sonenberg, and Frank Vetere.
\newblock Explainable reinforcement learning through a causal lens.
\newblock In {\em Proceedings of the AAAI Conference on Artificial
  Intelligence}, volume~34, pages 2493--2500, 2020.

\bibitem{miller_2021}
Tim Miller.
\newblock Contrastive explanation: a structural-model approach.
\newblock {\em The Knowledge Engineering Review}, 36:e14, 2021.

\bibitem{khan2009minimal}
Omar Khan, Pascal Poupart, and James Black.
\newblock Minimal sufficient explanations for factored markov decision
  processes.
\newblock In {\em Proceedings of the International Conference on Automated
  Planning and Scheduling}, volume~19, pages 194--200, 2009.

\bibitem{hayes2017improving}
Bradley Hayes and Julie~A Shah.
\newblock Improving robot controller transparency through autonomous policy
  explanation.
\newblock In {\em 2017 12th ACM/IEEE International Conference on Human-Robot
  Interaction (HRI}, pages 303--312. IEEE, 2017.

\bibitem{chakraborti2017plan}
Tathagata Chakraborti, Sarath Sreedharan, Yu~Zhang, and Subbarao Kambhampati.
\newblock Plan explanations as model reconciliation: Moving beyond explanation
  as soliloquy.
\newblock {\em arXiv preprint arXiv:1701.08317}, 2017.

\bibitem{sreedharan2019model}
Sarath Sreedharan, Alberto Olmo, Aditya~Prasad Mishra, and Subbarao
  Kambhampati.
\newblock Model-free model reconciliation.
\newblock {\em arXiv preprint arXiv:1903.07198}, 2019.

\bibitem{doshi2017towards}
Finale Doshi-Velez and Been Kim.
\newblock Towards a rigorous science of interpretable machine learning.
\newblock {\em arXiv preprint arXiv:1702.08608}, 2017.

\bibitem{booth2019evaluating}
Serena Booth, Christian Muise, and Julie Shah.
\newblock Evaluating the interpretability of the knowledge compilation map:
  Communicating logical statements effectively.
\newblock In {\em IJCAI}, pages 5801--5807, 2019.

\bibitem{tonekaboni2019clinicians}
Sana Tonekaboni, Shalmali Joshi, Melissa~D McCradden, and Anna Goldenberg.
\newblock What clinicians want: contextualizing explainable machine learning
  for clinical end use.
\newblock In {\em Machine learning for healthcare conference}, pages 359--380.
  PMLR, 2019.

\bibitem{hoffman2018metrics}
Robert~R Hoffman, Shane~T Mueller, Gary Klein, and Jordan Litman.
\newblock Metrics for explainable ai: Challenges and prospects.
\newblock {\em arXiv preprint arXiv:1812.04608}, 2018.

\bibitem{bansal2019updates}
Gagan Bansal, Besmira Nushi, Ece Kamar, Daniel~S Weld, Walter~S Lasecki, and
  Eric Horvitz.
\newblock Updates in human-ai teams: Understanding and addressing the
  performance/compatibility tradeoff.
\newblock In {\em Proceedings of the AAAI Conference on Artificial
  Intelligence}, volume~33, pages 2429--2437, 2019.

\bibitem{zhang2020effect}
Yunfeng Zhang, Q~Vera Liao, and Rachel~KE Bellamy.
\newblock Effect of confidence and explanation on accuracy and trust
  calibration in ai-assisted decision making.
\newblock In {\em Proceedings of the 2020 Conference on Fairness,
  Accountability, and Transparency}, pages 295--305, 2020.

\bibitem{kenny2021explaining}
Eoin~M Kenny, Courtney Ford, Molly Quinn, and Mark~T Keane.
\newblock Explaining black-box classifiers using post-hoc
  explanations-by-example: The effect of explanations and error-rates in xai
  user studies.
\newblock {\em Artificial Intelligence}, 294:103459, 2021.

\bibitem{cacioppo1984efficient}
John~T Cacioppo, Richard~E Petty, and Chuan Feng~Kao.
\newblock The efficient assessment of need for cognition.
\newblock {\em Journal of personality assessment}, 48(3):306--307, 1984.

\bibitem{goldberg1990alternative}
Lewis~R Goldberg.
\newblock An alternative" description of personality": the big-five factor
  structure.
\newblock {\em Journal of personality and social psychology}, 59(6):1216, 1990.

\bibitem{millecamp2019explain}
Martijn Millecamp, Sidra Naveed, Katrien Verbert, and J{\"u}rgen Ziegler.
\newblock To explain or not to explain: The effects of personal characteristics
  when explaining feature-based recommendations in different domains.
\newblock In {\em Proceedings of the 6th Joint Workshop on Interfaces and Human
  Decision Making for Recommender Systems}, volume 2450, pages 10--18. CEUR;
  http://ceur-ws. org/Vol-2450/paper2. pdf, 2019.

\bibitem{abbeel2004apprenticeship}
Pieter Abbeel and Andrew~Y Ng.
\newblock Apprenticeship learning via inverse reinforcement learning.
\newblock In {\em Proceedings of the twenty-first international conference on
  Machine learning}, page~1. ACM, 2004.

\bibitem{belanche2012integrating}
Daniel Belanche, Luis~V Casal{\'o}, and Carlos Flavi{\'a}n.
\newblock Integrating trust and personal values into the technology acceptance
  model: The case of e-government services adoption.
\newblock {\em Cuadernos de Econom{\'\i}a y Direcci{\'o}n de la Empresa},
  15(4):192--204, 2012.

\bibitem{graziano2012orientations}
William~G Graziano, Meara~M Habashi, Demetra Evangelou, and Ida Ngambeki.
\newblock Orientations and motivations: Are you a “people person,” a
  “thing person,” or both?
\newblock {\em Motivation and Emotion}, 36(4):465--477, 2012.

\bibitem{mayer2003three}
Richard~E Mayer and Laura~J Massa.
\newblock Three facets of visual and verbal learners: Cognitive ability,
  cognitive style, and learning preference.
\newblock {\em Journal of educational psychology}, 95(4):833, 2003.

\bibitem{dhanorkar2021needs}
Shipi Dhanorkar, Christine~T Wolf, Kun Qian, Anbang Xu, Lucian Popa, and Yunyao
  Li.
\newblock Who needs to know what, when?: Broadening the explainable ai (xai)
  design space by looking at explanations across the ai lifecycle.
\newblock In {\em Designing Interactive Systems Conference 2021}, pages
  1591--1602, 2021.

\end{thebibliography}

\appendix
\section{Explanation Types}
\label{sec:Explanation_examples}
This section shows the specific policy explanations shown to each participant for the four modalities employed in our study, i.e. (1) Basic Text, (2) Modified Text, (3) Decision Tree, (4) Program. 

\begin{figure}[!ht]
  \centering
  \subfloat[][]{\includegraphics[width=.4\textwidth]{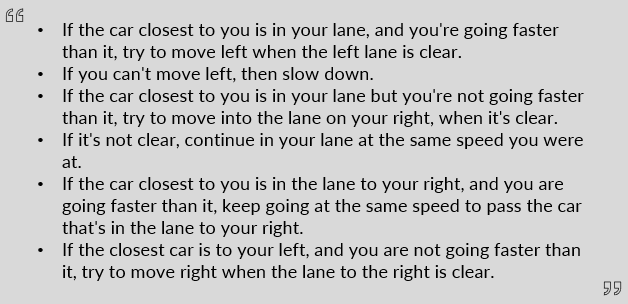}}\quad
  \subfloat[][]{\includegraphics[width=.4\textwidth]{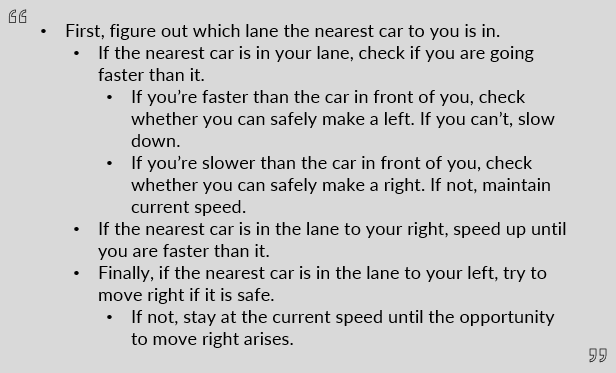}}\\
  \subfloat[][]{\includegraphics[width=.4\textwidth]{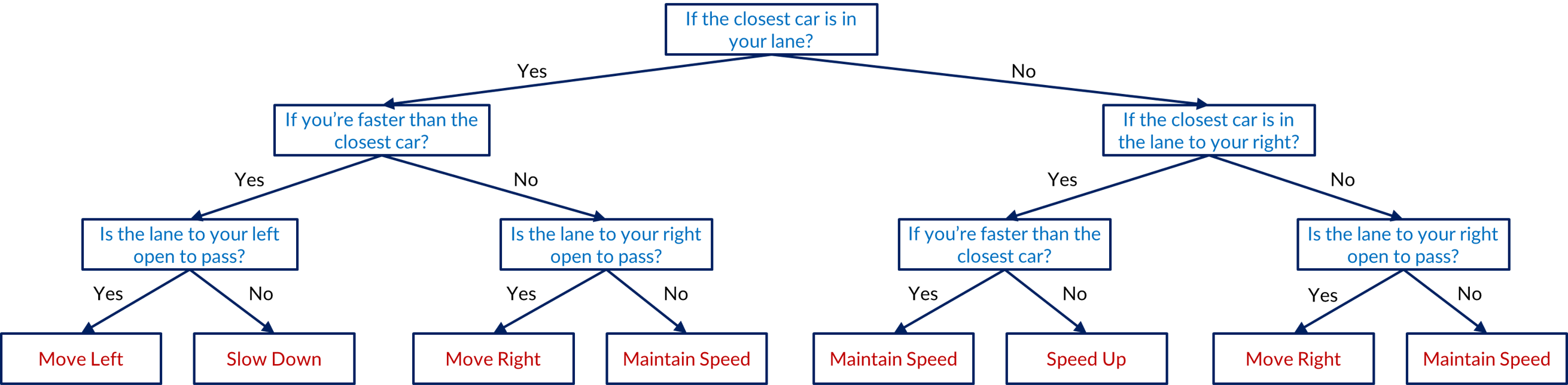}}\quad
  \subfloat[][]{\includegraphics[width=.4\textwidth]{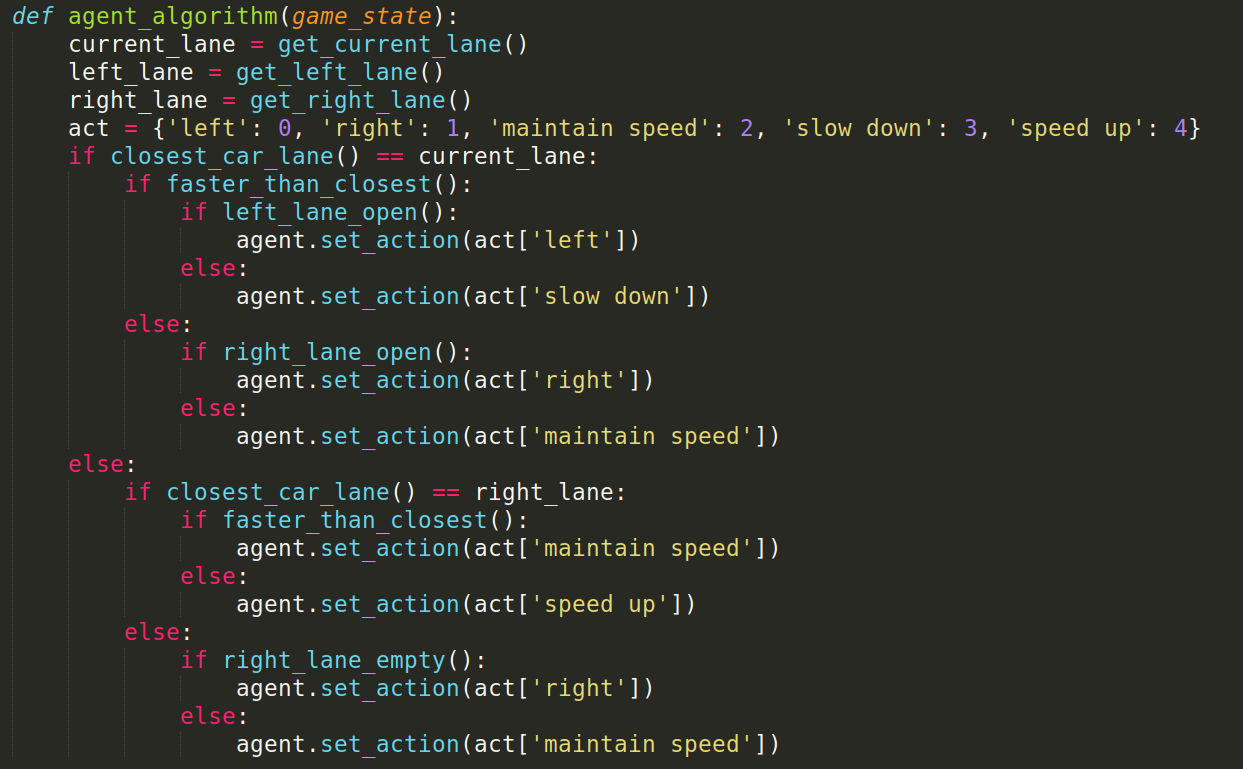}}
  \caption{This figure depicts the four policy explanations shown to participants corresponding to each baseline. (a) Basic Text: A language description generated using a template from the decision-tree policy, (b) Modified Text: A simplied version of the language description presented in an easy-to-understand manner, (c) Decision Tree: A decision tree describing the exact policy of the self-driving car, (d) Program: Pseudo-code of the decision making process of the car. }
  \label{fig:sub1}
\end{figure}

\end{document}